\documentclass[aps,showpacs,prb]{revtex4}
\usepackage{graphicx}
 
\newcommand{\be}{\begin{eqnarray}}
\newcommand{\ee}{\end{eqnarray}}
\newcommand{\bw}{\begin{widetext}}
\newcommand{\ew}{\end{widetext}}
\newcommand{\no}{\nonumber}
\newcommand{\Tr}{\mbox{Tr}\,}

\begin{document}
\title{Quantum fluctuations 
 in the transverse Ising spin glass model:  
 A field theory of random quantum spin systems}
\author{Kazutaka Takahashi}
\affiliation{Department of Physics, Tokyo Institute of Technology,  
 Tokyo 152--8551, Japan}
\date{\today}

\begin{abstract}
 We develop a mean-field theory for random quantum spin systems 
 using the spin coherent state path integral representation.
 After the model is reduced to the mean field one-body Hamiltonian, 
 the integral is analyzed 
 with the aid of several methods such as 
 the semiclassical method and the gauge transformation.
 As an application we consider the Sherrington-Kirkpatrick model
 in a transverse field.
 Using the Landau expansion and its improved versions, we give 
 a detailed analysis of the imaginary-time dependence of the order parameters.
 Integrating out the quantum part of the order parameters, 
 we obtain the effective renormalized free energy 
 written in terms of the classically defined order parameters.
 Our method allows us to obtain the spin glass-paramagnetic
 phase transition point $\Gamma/J\sim 1.62$ at $T=0$.
\end{abstract}
\pacs{
75.10.Nr, 
75.40.Gb, 
05.30.-d 
}
\maketitle

\section{Introduction}

 The classical random spin systems show various interesting properties 
 that cannot be observed in clean systems.~\cite{MPV,Nishimori}
 The main concern is the existence of the spin glass phase 
 and comprehensive analyses of several models such as 
 the Edwards-Anderson \cite{EA}
 and the Sherrington-Kirkpatrick (SK) model \cite{SK} 
 have revealed the properties of the randomness-induced phase transition.

 However, once we apply a transverse field to the SK model,
 the model becomes a quantum mechanical one and 
 cannot be solved exactly even at the mean field level.~\cite{CDS}
 Since the work of Bray and Moore \cite{BM} for the Heisenberg model,
 quantum spin glass models have attracted much interest as 
 they can study the interplay between randomness and quantum fluctuations.

 The effect of quantum fluctuations can be easily realized 
 by formulating the models in a path integral form.
 It is well known that 
 the $d$-dimensional quantum systems are mapped onto 
 the $(d+1)$-dimensional classical systems.~\cite{Sachdev}
 In the mean-field analysis, 
 models are described by 
 the functional integral of order parameters 
 and the additional coordinate (``time'') dependence 
 demonstrates the fluctuation effects.
 The simplest possible approximation is 
 to neglect the time dependence of order parameters, 
 which misses quantum effects and is not generally justified.

 Concerning the SK model in a transverse field, 
 the spin glass phase transition has been investigated 
 by many authors.~\cite{IY,U,YI,WL,Y,TLK,GL,UBK,MH,RGAR,KK}
 It was recognized there that the static approximation 
 is not justified at low temperature and 
 the spin-glass--paramagnetic-phase transition point at $T=0$
 is significantly affected by quantum effects.

 In this paper, we propose a systematic method 
 for quantum random spin systems. 
 It is based on expressing the partition function 
 using a path integral form.
 Using the Trotter-Suzuki decomposition \cite{TS} we insert 
 the spin coherent state representation as the resolution of unity.
 The use of coherent states has great advantage 
 for the resulting integrals 
 since the spin integral variables are continuous.
 Furthermore this method can be applied to the arbitrary Hamiltonian.
 The use of the spin coherent states is inevitable 
 for the quantum Heisenberg model and 
 was indeed done in Ref.\onlinecite{SYGPS}.
 Although it is not necessary for the SK model, 
 we stress in this paper that 
 our method is convenient for the systematic calculations.
 We mention, for example, 
 the semiclassical method and the gauge transformation.

 We apply our formulation to the transverse SK model.
 We carefully treat the time dependence of the order parameters 
 to investigate the spin-glass--paramagnetic-phase transition.
 We use several field theoretical methods: 
 the Landau expansion, 
 the renormalization-group-like method at finite temperature, 
 and the derivative expansion at zero temperature.
 The result of the phase diagram is compared to the previous works.

 The paper is organized as follows.
 In Sec.\ref{cs} we introduce a path integral form for 
 the general quantum spin Hamiltonian
 and discuss several methods to calculate the functional integral.
 In Sec.\ref{SKT} we apply the method to 
 the transverse SK model.
 After discussing the Landau expansion, 
 we consider improved methods to treat the quantum fluctuations.
 We calculate the spin glass order parameter 
 and determines the phase diagram.
 Section \ref{conc} is devoted to discussions and conclusions.

\section{Formulation for quantum spin systems}
\label{cs}

\subsection{Coherent state path integral representation}

 The path integral representation for spin systems 
 can be found in, e.g., Refs.\onlinecite{K,cs,WFS,AG}.
 The closure relation is expressed by using the spin coherent states 
 and is inserted to the Trotter-Suzuki decomposition \cite{TS} of 
 the matrix element of the time evolution operator.
 The spin operators in the Hamiltonian are replaced by 
 the classical variables, and 
 the complex phase factor is added to ensure the dynamics.

 The spin coherent state for spin $S$ is defined by 
\be
 |{\bf S} \rangle = 
 e^{-i\varphi\hat{S}_3} 
 e^{-i\theta\hat{S}_2} |S\rangle,
\ee
 where $\hat{{\bf S}}=(\hat{S}_1,\hat{S}_2,\hat{S}_3)$ 
 are spin operators satisfying
\be
 [\hat{S}_i,\hat{S}_j]=i\epsilon_{ijk}\hat{S}_k, \quad
 \hat{{\bf S}}^2 = S(S+1),
\ee
 and $|S\rangle$ denotes the eigenstate of $\hat{S}_3$ 
 with the eigenvalue $S$.
 $\theta$ and $\varphi$ are real parameters and 
 parametrize the coordinates on a unit sphere.
 The expectation value of the spin operators is given by 
\be
 {\bf S}=
 \langle{\bf S}|\hat{{\bf S}}|{\bf S}\rangle = 
 S(\sin\theta\cos\varphi, 
 \sin\theta\sin\varphi, \cos\theta). \label{s}
\ee
 The closure relation is written as 
\be
 \int d{\bf S}
 |{\bf S}\rangle\langle{\bf S}|=
 \frac{2S+1}{4\pi}\int_{-1}^1 
 d(\cos\theta) \int_0^{2\pi}d\varphi
 |{\bf S}\rangle\langle{\bf S}|=1.
\ee

 We consider the finite temperature partition function 
 for the quantum spin Hamiltonian $\hat{H}=H[\hat{{\bf S}}]$.
 We use the Trotter decomposition and insert 
 the coherent states defined above.
 We have the matrix element 
\be
 \langle {\bf S}(\tau+\Delta\tau)
 |e^{-\Delta\tau \hat{H}}|{\bf S}(\tau)\rangle
 \sim \exp\left\{\Delta\tau\left[-\langle {\bf S}(\tau)|
 \frac{d}{d\tau}|{\bf S}(\tau)\rangle
 -\langle {\bf S}(\tau)|\hat{H}|{\bf S}(\tau)\rangle
 \right]\right\},
\ee
 where $\tau$ is the slice index and represents the imaginary time
 between 0 and $\beta=1/k_BT$, and $\Delta \tau$ is 
 the slice width which must be taken $\Delta \tau\to 0$.
 The first term on the right-hand side is pure imaginary 
 and includes the time derivative.
 The second term gives the Hamiltonian with the operators 
 replaced by the classical variables given by Eq.(\ref{s}).
 Thus, we arrive at the expression for the finite temperature 
 partition function
\be
 Z = \int {\cal D}{\bf S} \exp\left\{
 i\int_0^\beta d\tau \Phi(\tau)
 -\int_0^\beta d\tau
 H[{\bf S}(\tau)]\right\}, \label{cspi}
\ee
 where
\be
 \Phi(\tau) = i\langle {\bf S}(\tau)|\frac{d}{d\tau}
 |{\bf S}(\tau)\rangle
 =S\dot{\varphi}(\tau)\cos\theta(\tau),
\ee
 and the periodic boundary condition 
 ${\bf S}(0)={\bf S}(\beta)$ is imposed.
 We note that $\Phi$ represents the geometric Berry phase.
 The corresponding term in Eq.(\ref{cspi}) is always imaginary
 irrespective of the real or imaginary time formulations 
 and describes the dynamical motion of ${\bf S}(\tau)$ on the sphere.

\subsection{Calculation methods}

 As we mentioned above, the advantage of the coherent state 
 representation is that the spin operators are replaced
 by the continuous classical variables, 
 which is crucial to describe the spin dynamics.
 Various methods developed in the path integral formalism 
 can be applied to the present case as well.
 Here we discuss two useful methods 
 used in the following sections.
 Without loss of generality in the present context 
 we can confine our discussion 
 to the one-body Hamiltonian
\be
 \hat{H} = -{\bf B}\cdot \hat{{\bf S}}, \label{BS}
\ee
 where $B$ is the external magnetic field 
 which may depend on time.
 In the following application of a spin glass model, 
 the system can be written in the one-body form by 
 introducing the auxiliary variables.

\subsubsection{Semiclassical method}

 First, we mention the semiclassical method which 
 is known as the standard approximation 
 in the path integral formalism. 
 We assume the main contribution comes from the stationary configuration
 of spin variables.
 The assumption is justified when $\hbar=0$ and 
 hence the name semiclassical.
 The stationary phase equation is nothing but 
 the classical equation of motion 
\be
 i\frac{d{\bf S}(\tau)}{d\tau}=
 {\bf S}(\tau)\times \frac{\partial H}{\partial {\bf S}(\tau)}.
 \label{eom}
\ee
 For the Hamiltonian (\ref{BS}), it is known that this approximation 
 becomes exact in the coherent state representation.
 The derived path integral representation is ill-defined and 
 we need a regularization to perform the integral explicitly.
 Klauder \cite{K} used the Wiener regularization and considered the 
 path integral using the stationary phase (saddle point) approximation.
 It was proved, after 20 years, that 
 the approximation becomes exact for the one-body Hamiltonian.~\cite{AG}
 The problem is reduced to solving  
 the classical equation of motion (\ref{eom})
 under the arbitrary boundary condition.
 The reason why the stationary phase approximation becomes exact 
 is that the coherent states satisfy the minimal uncertainty relation.

 In the following application discussed in the next section, 
 this method turns out to be useful 
 when we calculate the correlation function.
 For example, we consider the spin-$1/2$ system 
 with the Hamiltonian $\hat{H}=-2\Gamma \hat{S}_3$.
 The correlation function in the perpendicular (say, $y$) 
 direction to the magnetic field $\Gamma$
 can be written as
\be
 D(\tau-\tau') =  
 \frac{\int{\cal D}{\bf S} n_y(\tau)n_y(\tau')
 \exp\left[i\int_\tau \Phi(\tau)+\int_\tau \Gamma n_z(\tau)\right]}
 {\int{\cal D}{\bf S} 
 \exp\left[i\int_\tau \Phi(\tau)+\int_\tau \Gamma n_z(\tau)\right]},
 \label{Ddef}
\ee
 where ${\bf S}=S{\bf n}={\bf n}/2$ and $\int_\tau=\int_0^\beta d\tau$.
 The result of the integration is given by
\be
 D(\tau-\tau')=\frac{e^{\beta\Gamma-2\Gamma|\tau-\tau'|}
 +e^{-\beta\Gamma+2\Gamma|\tau-\tau'|}}
 {e^{\beta\Gamma}+e^{-\beta\Gamma}}. \label{D}
\ee
 This form is known and can be obtained by using other methods 
 such as the transfer matrix method.
 Competitive advantage of the present method arises when 
 we generalize the calculation to 
 higher order correlation functions and 
 higher spins with $S>1/2$.

\subsubsection{Gauge transformation}

 Second, the gauge transformation is utilized when 
 we calculate the correlation functions.
 The basis of the state to be inserted into the Trotter 
 decomposition can be changed to other arbitrary basis.
 We consider the rotation in spin space by the unitary operator
\be
 \hat{U}[\mbox{{\boldmath $\phi$}}(\tau)]=\exp\left[i
 \int_\tau \mbox{{\boldmath $\phi$}}(\tau)\cdot\hat{\bf S}\right].
\ee
 The partition function of 
 the Hamiltonian (\ref{BS}) can be written as 
\be
 Z = \int {\cal D}{\bf S} \exp\left[
 i\int_\tau\Phi(\tau)
 +\int_\tau 
 {\bf B}(\tau)\cdot {\bf S}(\tau)
 \right].
\ee
 We consider the time-dependent rotation 
 diagonalizing the Hamiltonian to 
 $\hat{H}=-|{\bf B}(\tau)|\hat{S}_3$.
 Such a choice is always possible, but
 this does not solve the problem 
 because we have the expression 
\be
 Z = \int {\cal D}{\bf S} \exp\left\{
 i\int_\tau\left[\Phi(\tau)
 +\dot{\mbox{{\boldmath $\phi$}}}(\tau)\cdot{\bf S}(\tau)\right]
 +\int_\tau 
 |{\bf B}(\tau)|S_3(\tau)
 \right\}.
 \label{gp}
\ee
 Due to the time-dependent gauge transformation, 
 the phase acquires an extra term.
 This cannot be solved generally except simple cases 
 such as constant fields and oscillating fields.
 This type of the gauge transformation was 
 discussed in Ref.\onlinecite{SF} and 
 the extra term of Eq.(\ref{gp}) is called the geometric phase.
 This expression can be utilized when 
 we consider the derivative expansion.
 When the time-dependence is slow, the geometric phase term is 
 expanded and the only remaining thing to do is to 
 calculate the correlation functions as Eq.(\ref{Ddef}).

\section{Application to the Sherrington-Kirkpatrick model 
 in a transverse field}
\label{SKT}

 As an application of the spin coherent path integral 
 representation we consider the transverse SK model 
 defined by the Hamiltonian with Pauli spins 
 $\hat {\bf S}=\mbox{\boldmath $\sigma$}/2$ 
 on lattice sites
\be
 \hat{H} = -\frac{1}{2}\sum_{i\ne j}^N J_{ij}
 \sigma^z_i \sigma^z_j
 -\Gamma\sum_{i=1}^N \sigma_i^x,
\ee
 where $J_{ij}$ is the Gaussian random variables with mean $J_0/N$
 and variance $J^2/N$, $N$ is the number of lattice sites, 
 and $\Gamma$ is the transverse field.
 $i$ and $j$ run over all points, 
 which means that the interaction is infinite range 
 and the mean-field analysis becomes exact.
 When the transverse field is present, the model cannot be solved 
 exactly and we need an approximation.
 As we explained in the Introduction, 
 the simplest way to proceed is to neglect  
 the time dependence of the order parameters 
 introduced as auxiliary integral variables.
 It looks plausible because the time variable is introduced artificially 
 and its dependence cannot directly be observed.
 However, we discuss in the following that 
 the time dependence should be integrated out rather than be neglected,
 which gives a nontrivial quantum effect.

\subsection{Mean field theory}

 We follow the standard prescription to introduce 
 the auxiliary fields $m$ and $q$ using the Hubbard-Stratonovich
 transformation.~\cite{MPV,Nishimori}
 Introducing replicas, we take 
 the average of the $n$th power of the partition function as 
\be
 [Z^n] &=& 
 \int {\cal D}{\bf S} \exp\biggl\{
 \sum_{a=1}^n\sum_{i=1}^{N}\int_\tau 
 \left[i\Phi_{i}^a(\tau)+\Gamma n_{xi}^a(\tau)\right]
 \no\\
 & & +\frac{J_0}{N}\sum_{a=1}^n\sum_{i<j}^N\int_\tau
  n^a_{zi}(\tau) n^a_{zj}(\tau)
 +\frac{J^2}{2N}\sum_{a,b=1}^{n}\sum_{i<j}^N
 \int_{\tau,\tau'}
 n^a_{zi}(\tau) n^a_{zj}(\tau)n^b_{zi}(\tau') n^b_{zj}(\tau')
 \biggr\}.
\ee
 The Hubbard-Stratonovich transformation allows us to 
 introduce the order parameters.
 We have
\be
 [Z^n] = 
 \int {\cal D}m{\cal D}q
 \exp\biggl\{
 -\frac{NJ_0}{2}\sum_{a=1}^n\int_\tau \left[m_a(\tau)\right]^2
 -\frac{NJ^2}{4}\sum_{a,b=1}^n\int_{\tau,\tau'}
 \left[q_{ab}(\tau,\tau')\right]^2
 +N\ln\Tr e^L\biggr\},
\ee
 where
\be
 \Tr e^L &=&
 \int {\cal D}{\bf S} \exp\biggl\{
 \sum_{a=1}^n\int_\tau 
 \left[i\Phi^a(\tau)+\Gamma n_{x}^a(\tau)
 +J_0m_a(\tau)n^a_{z}(\tau)
 \right]
 +\frac{J^2}{2}\sum_{a,b=1}^n\int_{\tau,\tau'}
 q_{ab}(\tau,\tau')n^a_{z}(\tau)n^b_{z}(\tau')
 \biggr\}.
\ee
 The saddle point equations
\be
 m_a(\tau) = \frac{\Tr n_z^a(\tau)e^L}{\Tr e^L}, \quad
 q_{ab}(\tau,\tau') = 
 \frac{\Tr n_z^a(\tau)n_z^b(\tau')e^L}{\Tr e^L}
\ee
 indicate that $m$ is the magnetization and
 $q$ the spin glass order parameter.

\subsection{Landau expansion}

 To proceed further, we must integrate out the spin variables 
 ${\bf n}^a(\tau)$ to get the order parameter functional.
 We do this by using the Landau expansion assuming
 the order parameters are small.
 The Landau theory for the present model was considered 
 in Ref.\onlinecite{RSY} 
 by writing down the functional immediately from the symmetry argument.
 That method is phenomenological 
 and the coupling constants in the Landau function 
 cannot be related to 
 the fundamental parameters in the original Hamiltonian.
 Here we derive the Landau function microscopically from the original Hamiltonian. 
 We use the derived result to determine the phase boundary 
 where the approximation makes sense.
 Then the transition point can be expressed by $T/J$ and $\Gamma/J$.

 In order to get the result we must calculate the spin correlation 
 function (\ref{D}).
 Higher order correlation functions can be calculated in the same way
 as the two point function.
 For instance the four point function is 
\be
 D(\tau_1,\tau_2,\tau_3,\tau_4) = D(\tau_1'-\tau_2'+\tau_3'-\tau_4'),
\ee 
 where $\tau_{1,2,3,4}$ are arranged in order of magnitude as 
 $\tau_1'>\tau_2'>\tau_3>\tau_4'$.
 Higher order functions can be expressed in the same way.

 Now we can express each term of the Landau expansion 
 using the correlation functions.
 For simplicity we consider $J_0m_a(\tau)=0$ which means we consider the
 paramagnetic or spin glass phase.
 We write for the spin glass order parameter
\be
 q_{ab}(\tau,\tau')=\delta_{ab}\chi_a(\tau,\tau')
 +(1-\delta_{ab})q_{ab}(\tau,\tau'),
\ee
 to distinguish the role of each term.
 The first term is the diagonal part in the replica space
 and represents the spin susceptibility.
 It is unity when $\Gamma=0$ and 
 the deviation from unity at $\Gamma\ne 0$ 
 represents the quantum effect.
 The second term is the familiar spin glass order parameter.
 
 Using this representation, 
 we can Landau-expand the averaged free energy as
\be
 \beta[f] = -\ln(e^{\beta\Gamma}+e^{-\beta\Gamma})
 +\lim_{n\to 0}\frac{1}{n}\left\{
 \frac{J^2}{4}\sum_{a=1}^n\int_{\tau} \chi_a^2(\tau)
 +\frac{J^2}{4}\sum_{a\ne b=1}^n\int_{\tau\tau'}q_{ab}^2(\tau,\tau')
 -\sum_{k=1}^\infty\frac{1}{k!}\left(\frac{J^2}{2}\right)^kI_k
 \right\},
\ee
 where $I_k$ is of $k$th order in $\chi$ and $q$.
 For $k=1$ and 2, it is given by
\be
 && I_1 = \sum_{a=1}^n\int_{\tau_1\tau_2} 
 \chi_{a}(\tau_1,\tau_2)D(\tau_1-\tau_2), \\
 && I_2 = 
 \sum_{a=1}^n\int_{\tau_1\tau_2\tau_3\tau_4} 
 \chi_{a}(\tau_1,\tau_2)\chi_{a}(\tau_3,\tau_4)
 \left[D(\tau_1,\tau_2,\tau_3,\tau_4)
 -D(\tau_1-\tau_2)D(\tau_3-\tau_4)\right] \no\\
 && +2\sum_{a\ne b}^n\int_{\tau_1\tau_2\tau_3\tau_4} 
 q_{ab}(\tau_1,\tau_3)q_{ba}(\tau_4,\tau_2)
 D(\tau_1-\tau_2)D(\tau_3-\tau_4).
\ee
 We note that this is the expansion in terms of 
 $\beta^2J^2q_{ab}(\tau,\tau')$ and $\beta^2J^2\chi_{a}(\tau,\tau')$.
 Since the order parameters are unity at most, 
 we can regard $\beta^2J^2$ as a formal expansion parameter. 
 The order parameters are determined by the saddle point equations.
 In the following we consider 
 the replica symmetric and nonsymmetric cases.
 
\subsubsection{Replica symmetric solution}

 We assume that $\chi$ and $q$ are independent of the replica index.
 Then the saddle point equations up to first order in $\beta^2J^2$
 are given by 
\be
 && \chi(\tau_1,\tau_2)=D(\tau_1-\tau_2)+\frac{J^2}{2}
 \int_{\tau_3\tau_4}\chi(\tau_3,\tau_4)
 \left[D(\tau_1,\tau_2,\tau_3,\tau_4)
 -D(\tau_1-\tau_2)D(\tau_3-\tau_4)\right],  
 \label{chilandau} \\
 && q(\tau_1,\tau_2) = J^2\int_{\tau_3\tau_4}
 q(\tau_3,\tau_4)D(\tau_1-\tau_3)D(\tau_4-\tau_2). 
 \label{qlandau}
\ee
 At the linear approximation $\chi(\tau,\tau')$ 
 is equal to the correlation function $D(\tau-\tau')$
 and the assumption of the replica independence is plausible 
 in the perturbative calculation.
 On the other hand, the time dependence of $\chi$ cannot be neglected.
 Concerning $q$, the static approximation seems to be appropriate
 as we discuss below.

 $\chi$ can be solved iteratively while 
 $q$ is solved by considering higher order nonlinear terms.
 The phase boundary can be determined by 
 the leading term in Eq.(\ref{qlandau}).
 Using the static approximation for $q$, we obtain 
\be
 1 =J^2\left(\int_{\tau}D(\tau)\right)^2
 = \beta^2J^2\left(\frac{\tanh\beta\Gamma}{\beta\Gamma}\right)^2.
 \label{qblandau}
\ee
 This can be easily solved to find $\Gamma/J=1$ at $T=0$.~\cite{WL}
 This perturbative solution is compared to the static approximation 
 result $\Gamma/J=2$ in Refs.\onlinecite{U} and \onlinecite{TLK}
 where $\chi$ is not expanded and treated nonperturbatively.
 The difference comes from the fact that 
 $\chi$ does not vanish at the phase boundary and 
 contributes to Eq.(\ref{qblandau}).
 
\subsubsection{Replica symmetry breaking solution}

 It is well known in the classical SK model without transverse field 
 that the spin glass order parameter depends on the replica index and 
 the replica symmetry breaking (RSB) solution 
 proposed by Parisi \cite{Parisi} is the exact one.
 Here we consider the effect of the transverse field for the RSB solution.
 The calculation can be done explicitly if we use the Landau expansion.
 The free energy is expanded in $q_{ab}$ up to the fourth order and 
 the saddle-point equation is solved analytically 
 under the assumption of the RSB.~\cite{Parisi,Nishimori}
 This can be done even if the transverse field is incorporated. 
 Near the transition point $T=T_c$, we obtain the expression
\be
 \beta [f] &=& \lim_{n\to 0}\frac{1}{n}\biggl\{
 \frac{1}{2}\theta\sum_{a\ne b}(Q_{ab})^2
 -\frac{1}{6}C_1\sum_{a\ne b\ne c}Q_{ab}Q_{bc}Q_{ca}
 -\frac{1}{12}C_{2}\sum_{a\ne b}(Q_{ab})^4 \no\\
 & & +\frac{1}{4}C_3\sum_{abc}(Q_{ab})^2(Q_{ac})^2
 -\frac{1}{8}C_4\Tr(Q)^4
 \biggr\}, \label{lexp}
\ee
 where $Q_{ab}=\beta^2 J^2 q_{ab}$ and $\theta=(T_c-T)/T_c$.
 $T_c$ and the coefficients $C_{1,2,3,4}$ depend on $\chi$ and 
 are expressed perturbatively.
 We have for $C_{1,2}$
\be
 & & C_1 = \left(\frac{1}{\beta}\int_{\tau}D(\tau)\right)^3
 +\frac{3}{2}\beta^2J^2\left(\frac{1}{\beta}\int_{\tau}D(\tau)\right)^2
 \frac{1}{\beta^4}\int_{\tau_{1-4}}
 \chi(\tau_1,\tau_2)\left[D(\tau_1,\tau_2,\tau_3,\tau_4)
 -D(\tau_1-\tau_2)D(\tau_3-\tau_4)\right]+\cdots, \no\\
 & & C_2 = -\frac{1}{2\beta^8}\int_{\tau_{1-8}}
 \left[
 D(\tau_1,\tau_2,\tau_3,\tau_4)D(\tau_5,\tau_6,\tau_7,\tau_8)
 -3D(\tau_1-\tau_2)D(\tau_3-\tau_4)D(\tau_5-\tau_6)D(\tau_7-\tau_8)
 \right]+\cdots.
 \label{C12}
\ee
 $T_c$ and $C_{3,4}$ are expressed in a similar way.
 When $\Gamma=0$, all the correlation functions are set to unity
 and we obtain  the classical result $C_{1,2,3,4}=1$.
 Since the effect of the transverse field is only 
 to change the coefficients of the expansion (\ref{lexp}),
 the saddle point equation for $Q_{ab}$ 
 is easily solved in the same way as the classical case as
\be
 && q(x)=\frac{C_1}{C_{2}}\frac{x}{2} \quad 
 (0\le x < x_1=2|\theta| C_2/C_1), \no\\
 && q(x)=|\theta| \quad (x_1\le x \le 1),
\ee
 where $x$ is the replica continuous variable at $n\to 0$.
 Thus, the transverse field changes the slope of the line.
 Each term of the coefficient $C_{1,2}$ can be calculated analytically.
 If we keep only the leading term in Eq.(\ref{C12})
 we have 
\be
 \frac{C_1}{C_2} \sim 
 \left(\frac{\tanh\beta\Gamma}{\beta\Gamma}\right)^3
 \frac{1}{\frac{3}{2}\left(\frac{\tanh\beta\Gamma}{\beta\Gamma}\right)^4
 -\frac{9}{2(\beta\Gamma)^4}
 \left(1-\frac{\tanh\beta\Gamma}{\beta\Gamma}\right)^2}.
\ee
 This is larger than unity when $\beta\Gamma>0$, 
 which means that the RSB solution approaches 
 the replica symmetric solution
 and we expect that 
 the stability of the replica symmetric solution increases.
 In the following calculations, 
 we consider the replica symmetric solution 
 and our attention is mainly focused on 
 the time dependence of $\chi$.

\subsection{Improved Landau expansion at classical regime}

 As we mentioned above 
 the naive perturbative expansion 
 gives the transition point $\Gamma=J$ at $T=0$ (Ref.\onlinecite{WL}) and 
 the static approximation for $\chi$ gives 
 $\Gamma=2J$ at $T=0$.~\cite{U,TLK}
 Furthermore, several analyses using more sophisticated techniques
 showed that the transition point lies between them.~\cite{YI,MH}
 In the following analysis, we reconsider this problem using 
 a refined field theoretical method systematically.
 Our method allows us to obtain 
 the phase structure not only at the boundary 
 as was done in previous works 
 but also in the whole space.

 Equation (\ref{chilandau}) tells us that 
 the order parameter $\chi(\tau_1,\tau_2)$ is approximately equal to 
 the correlation function $D(\tau_1-\tau_2)$ in Eq.(\ref{D}).
 This function has a slow time dependence at $\beta\Gamma\ll 1$
 and fast at $\beta\Gamma\gg 1$.
 Therefore, when the temperature is not so low, 
 the static approximation is expected to be valid.
 On the other hand, 
 it is not justified at low temperature.

 First we treat the classical regime where the time dependence is 
 not so strong.
 In this case we can separate 
 the time-dependent and -independent parts of 
 $\chi_a(\tau)$ and $q_{ab}(\tau,\tau')$ as
\be
 \chi_a(\tau)=\chi+\tilde{\chi}_a(\tau), \quad
 q_{ab}(\tau,\tau')=q+\tilde{q}_{ab}(\tau,\tau'),
\ee
 where $\chi$ and $q$ are zero modes defined as 
 the zero frequency part of the Fourier transformation.
 $\tilde{\chi}_a(\tau)$ and $\tilde{q}_{ab}(\tau,\tau')$
 are expected to be small and we consider the Landau expansion
 in terms of these variables.
 On the other hand $\chi$ and $q$ are not expanded and 
 treated nonperturbatively
 as was done in the static calculation.
 We expand the averaged free energy 
 in terms of the nonstatic modes
 and integrate those modes as
\be
 [Z^n]= \int {\cal D}q{\cal D}\chi
 {\cal D}\tilde{q}{\cal D}\tilde{\chi}
 \exp\left\{-Nn\beta
 f[q,\chi,\tilde{q},\tilde{\chi}]\right\} 
 = \int {\cal D}q{\cal D}\chi
 \exp\left\{-Nn\beta
 f_{\rm eff}[q,\chi]\right\}.
\ee
 $f_{\rm eff}$ is defined as the classical free energy 
 renormalized by the quantum part of the order parameters.
 $f$ is expanded in $\tilde{\chi}$ and $\tilde{q}$
 up to second order and 
 we carry out the Gaussian integrals.
 Thus the static and nonstatic parts are treated separately and 
 we can take full advantage of both the Landau expansion 
 and the static approximation.

 Using the separation of order parameters, we have 
\be
 \Tr e^L &=&
 \int {\cal D}{\bf S} Dz \prod_{a}Dz_a \exp\biggl[
 \sum_{a}\int_\tau 
 \left[i\Phi^a(\tau)+\Gamma n_{x}^a(\tau)
 +J(\sqrt{q}z+\sqrt{\chi-q}z_a)n_{z}^a(\tau)
 \right]
 \no\\
 & & \qquad\qquad
 +\frac{J^2}{2}\sum_{a}\int_{\tau\tau'}
 \tilde{\chi}_{a}(\tau,\tau')n^a_{z}(\tau) n^a_{z}(\tau')
 +\frac{J^2}{2}\sum_{a\ne b}\int_{\tau\tau'}
 \tilde{q}_{ab}(\tau,\tau')n^a_{z}(\tau) n^b_{z}(\tau')
 \biggr],
 \label{trel}
\ee
 where we introduced the auxiliary variables $z$ and $z_a$
 ($a=1,2,\ldots, n$), 
 and the integration measures are given by 
\be
 Dz = \frac{dz}{\sqrt{2\pi}} e^{-z^2/2}, \quad 
 Dz_a = \frac{dz_a}{\sqrt{2\pi}} e^{-z_a^2/2}.
 \label{Dz}
\ee
 The last two terms are expanded up to second order as
\be
 \Tr e^L &=&
 \int Dz \prod_{a=1}^nDz_a 
 \prod_{a=1}^n (e^{\beta h_a}+e^{-\beta h_a})
 \exp\biggl[
 \frac{J^2}{2}\sum_{a=1}^n \frac{\Gamma^2}{h_a^2}\int d\tau d\tau' 
 \tilde{\chi}_a(\tau,\tau')D_{h_a}(\tau-\tau') 
 \no\\
 & & +\frac{J^4}{8}\sum_a\int_{\tau_{1,2,3,4}}
 \tilde{\chi}_a(\tau_1,\tau_2)
 \tilde{\chi}_a(\tau_3,\tau_4)\left[
 D_{h_a}(\tau_1,\tau_2,\tau_3,\tau_4)
 -D_{h_a}(\tau_1,\tau_2)D_{h_a}(\tau_3,\tau_4)\right]
 \no\\
 & & +\frac{J^4}{4}\sum_{a\ne b}\int_{\tau_{1,2,3,4}}
 \tilde{q}_{ab}(\tau_1,\tau_2)
 \tilde{q}_{ab}(\tau_3,\tau_4)
 D_{h_a}(\tau_1,\tau_3)D_{h_b}(\tau_2,\tau_4)
 \frac{\Gamma^2}{h_a^2}\frac{\Gamma^2}{h_b^2}
 \biggr],
\ee
 where 
\be
 h_a = \sqrt{\Gamma^2+J^2(\sqrt{q}z+\sqrt{\chi-q}z_a)^2},
\ee
 and $D_{h_a}(\tau,\tau')$ is the correlation function (\ref{D})
 with $\Gamma$ replaced by $h_a$.
 We used the gauge transformation to ``diagonalize'' the Hamiltonian.
 In the present case, the magnetic field is independent of time and 
 the extra phase factor does not arise here.
 Then taking the $n\to 0$ limit we obtain 
\be
 \lim_{n\to 0} \ln\Tr e^L 
 &=& \int Dz_1 \ln\left[\int Dz_2 (e^{\beta h}+e^{-\beta h})\right] 
 \no\\ & & 
 +\lim_{n\to 0}\frac{1}{n}\biggl[
 \frac{J^2}{2}\sum_{a=1}^n\int_{\tau\tau'}
 \tilde{\chi}_a(\tau,\tau')\int Dz_1D'z_2 
 D_h(\tau-\tau')\frac{\Gamma^2}{h^2} \no\\
 & &
 +\frac{J^4}{8}\sum_a\int_{\tau_{1,2,3,4}}
 \tilde{\chi}_a(\tau_1,\tau_2)\tilde{\chi}_a(\tau_3,\tau_4)
 \int Dz_1D'z_2 
 \left[D_h(\tau_1,\tau_2,\tau_3,\tau_4)
 -D_h(\tau_1-\tau_2)D_h(\tau_3-\tau_4)\right]
 \no\\
 & &
 +\frac{J^4}{4}\sum_{a\ne b}\int_{\tau_{1,2,3,4}}
 \tilde{q}_{ab}(\tau_1,\tau_2)\tilde{q}_{ab}(\tau_3,\tau_4)
 \int Dz_1
 \left(\int D'z_2 D_h(\tau_1-\tau_3)\frac{\Gamma^2}{h^2}\right)
 \left(\int D'z_2 D_h(\tau_2-\tau_4)\frac{\Gamma^2}{h^2}\right)
 \biggr], \no\\
 \label{tildeexp}
\ee
 where $h$ is equal to $h_a$ with $z$ ($z_a$) replaced by $z_1$ ($z_2$), and 
\be
 Dz_{1,2} =
 \frac{dz_{1,2}}{\sqrt{2\pi}}
 e^{-z_{1,2}^2/2}, \qquad
 \int D'z_2 \left(\cdots\right)=
 \frac{\int Dz_2\left(\cdots\right)\cosh\beta h}{\int Dz_2\cosh\beta h}.
\ee
 The integrations of the nonzero modes 
 $\tilde{\chi}_a(\tau)$ and $\tilde{q}_{ab}(\tau)$ 
 are easily carried out to obtain 
 the renormalized effective free energy
 with the leading nontrivial contribution 
\be
 \beta f_{\rm eff} \sim \frac{\beta^2 J^2}{4}(\chi^2-q^2)
 -\int Dz_1\ln\left[\int Dz_2 (e^{\beta h}+e^{-\beta h})\right]
 -\frac{\beta^2 J^2}{2}\sum_{m=1}^\infty
 \left(\int Dz_1D'z_2 \tilde{D}_h(m)\frac{\Gamma^2}{h^2}\right)^2,
 \label{feff}
\ee
 where $\tilde{D}_h(m)$ 
 is the Fourier transformation of $D_h(\tau)$ and is given by
\be
 \tilde{D}_h(m) = \frac{(\beta h)^2}{(\beta h)^2+(\pi m)^2}
 \frac{\tanh\beta h}{\beta h}.
\ee
 Up to the second order expansion of the nonzero modes,
 $\tilde{q}$ fluctuations do not contribute to the result.
 This is because the spin glass order parameter is defined 
 as the global variable in terms of the lattice site index 
 as $F=Nf[q]$.
 A different conclusion is obtained 
 if we define $q$ as local order parameters $q_i$ 
 and write $F[q]=\sum_i f[q_i]$.
 Fluctuations in each local site give a nontrivial result.
 Our model was defined as the infinite range model.
 In that case, $\tilde{q}$ fluctuations are not important 
 and the static approximation for $\tilde{q}$ is justified.
 In this sense the mean field theory of the infinite range
 model is different from that of the finite range model as discussed
 in Ref.\onlinecite{RSY}.

\begin{figure}[htb]
\begin{center}
\includegraphics[width=0.5\columnwidth]{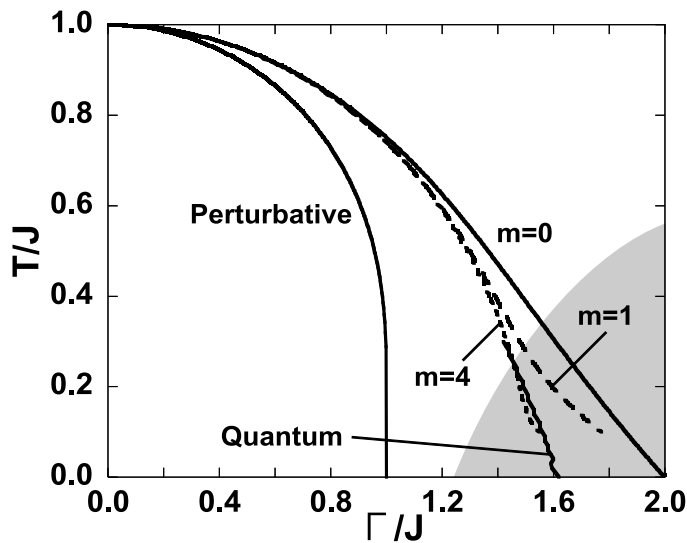}
\caption{Phase boundary from the saddle point equations (\ref{speqb}). 
 The maximum frequency value of the summation is denoted with lines.
 ``$m=0$'' corresponds to the static approximation result and
 ``Perturbative'' denotes the naive Landau expansion result 
 (\ref{qblandau}).
 ``Quantum'' is the result in Sec.\ref{qr}.
 The quantum regime is shown by the shaded area.}
\label{spb}
\end{center}
\end{figure}

 The effective free energy (\ref{feff}) is a function of $q$ and $\chi$.
 Their values are obtained by solving the saddle point equations
 $\partial f_{\rm eff}/\partial\chi=0$ and
 $\partial f_{\rm eff}/\partial q=0$.
 The phase boundary is determined by 
 $\partial f_{\rm eff}/\partial\chi|_{q=0}=0$ and
 $\partial^2 f_{\rm eff}/\partial^2 q|_{q=0}=0$.
 Using formulas derived in the Appendix, we obtain 
\be
 & & \chi = \frac{1}{\beta^2J^2\chi}\left(\int D'z z^2-1\right)
 +\frac{1}{\chi}\sum_{m=1}^\infty
 \left(\int D'z \tilde{D}_h(m)\frac{\Gamma^2}{h^2}\right)
 \left[\int D'z \left(z^2-\int D'z' z'^2\right)
 \tilde{D}_h(m)\frac{\Gamma^2}{h^2}\right], \no\\
 & &
 (\beta J\chi)^2=\frac{\left(\int D'z z^2-1\right)^2}{2\int D'z z^2-3},
 \label{speqb}
\ee
 where $h=\sqrt{\Gamma^2+J^2\chi z^2}$.

\begin{figure}[htb]
\begin{center}
\includegraphics[width=0.5\columnwidth]{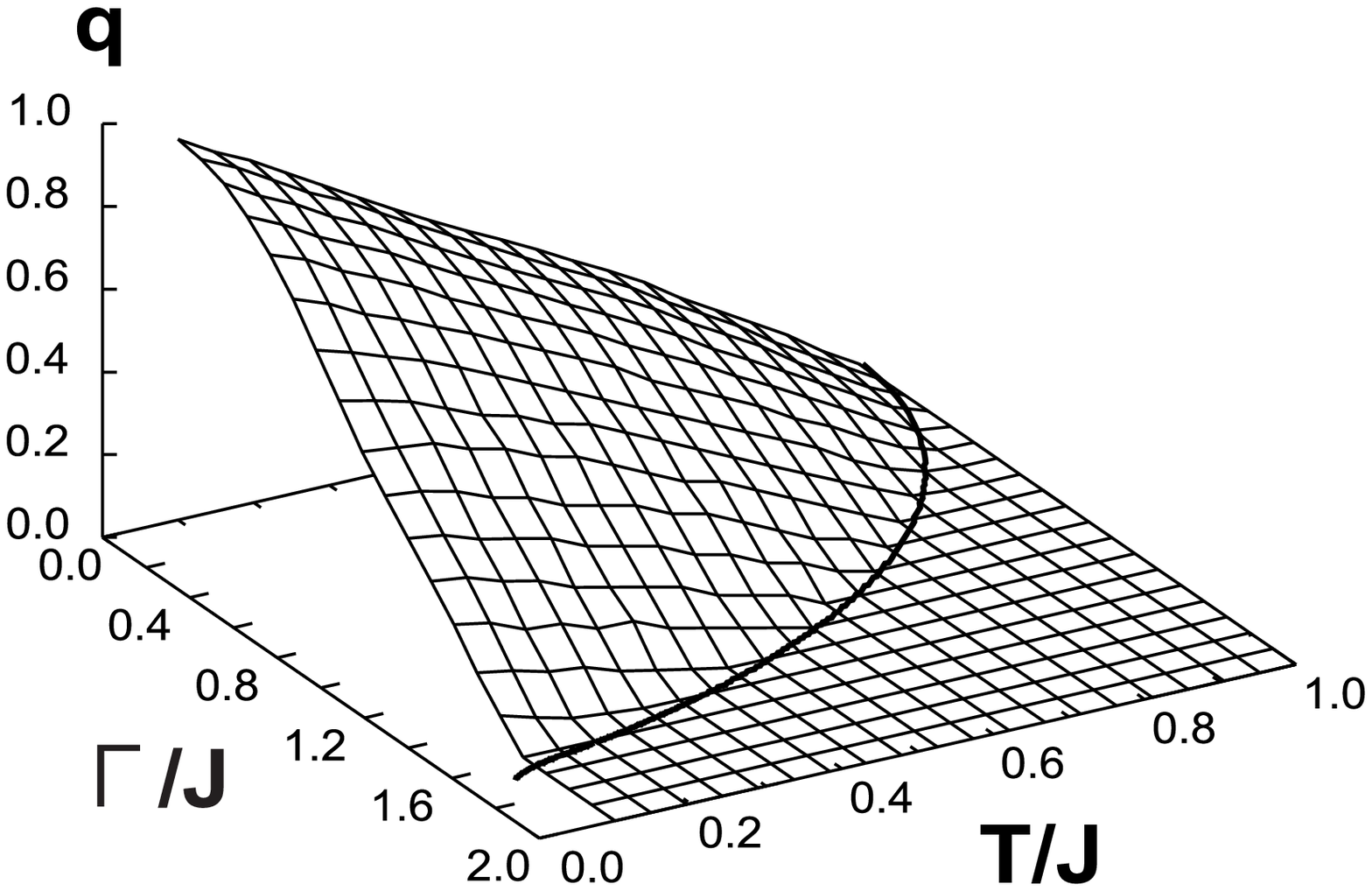}
\includegraphics[width=0.5\columnwidth]{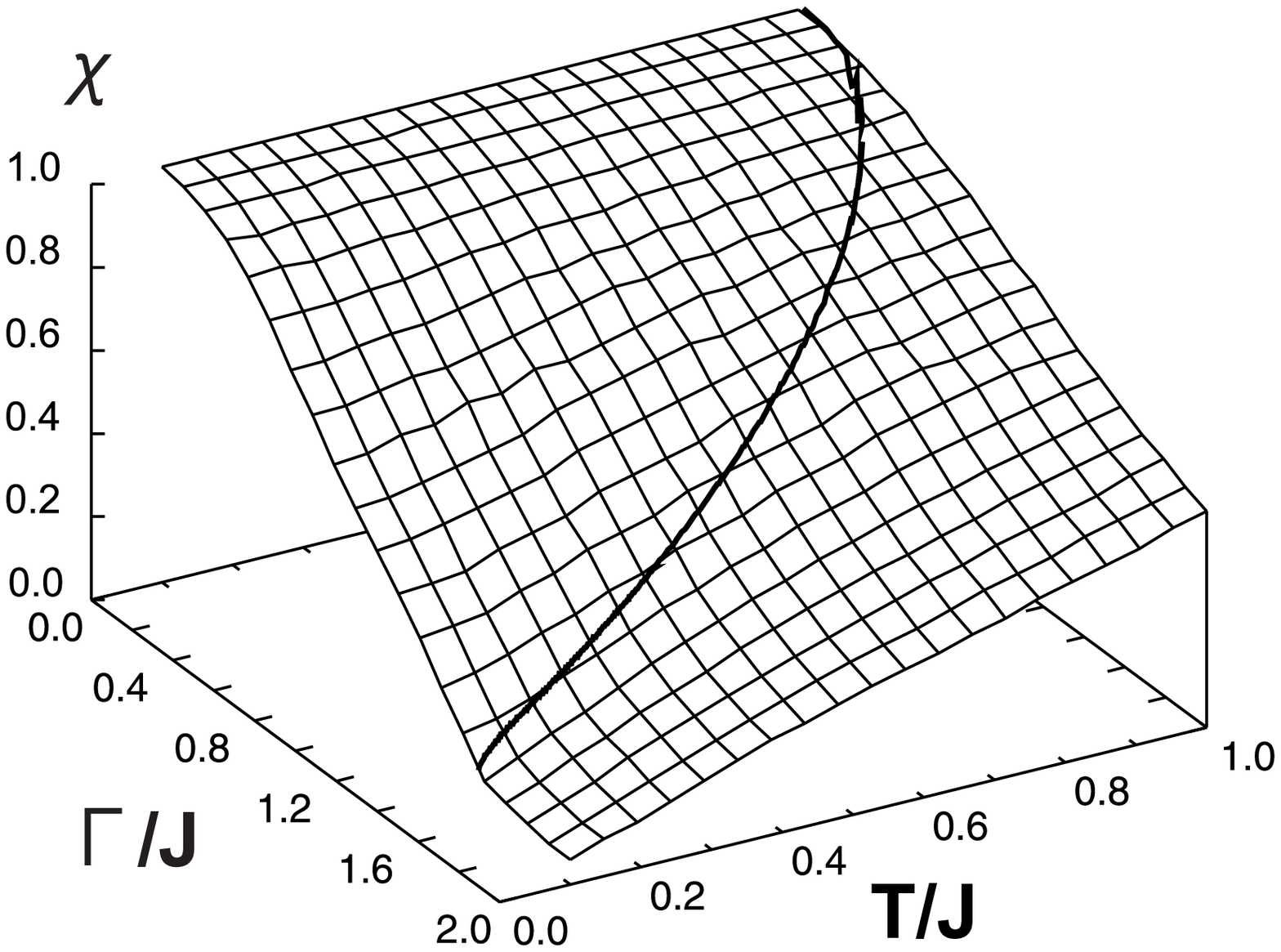}
\caption{Spin glass order parameters $q$ and $\chi$ 
 from saddle point equations.
 The frequency summation is taken up to $m=4$. 
 The bold line denotes the phase boundary.
}
\label{qchi}
\end{center}
\end{figure}

 We show the numerically solved result in Fig.\ref{spb}.
 The frequency summation is restricted to a finite value and 
 we denote in the figure the maximum value of the summation.
 The line $m=0$ corresponds to the static approximation result
 shown before
 and ``perturbative'' means the Landau expansion result (\ref{qblandau}).
 We see up to $T/J \sim 0.2$ from the above 
 the maximum value $m=4$ is sufficient to find the convergence.
 The equations cannot be solved at lower temperature.
 Actually we can show analytically that the equations (\ref{speqb})
 do not have a real solution at $T=0$.
 The extrapolation of the finite-$T$ results gives 
 $\Gamma/J\sim 1.54$ at $T\to 0$, which is close to 
 the results obtained in previous works.~\cite{YI,MH,RGAR}
 We also show the numerical result of the order parameters 
 in Fig.\ref{qchi}.
 The behavior of $\chi$ shows that the quantum effect 
 becomes important at low temperatures and large transverse fields 
 as expected.
 We also see $\chi\sim q$ at low temperature,
 which implies the reduction of the effective number of 
 the order parameters.

\subsection{Improved Landau expansion at quantum regime}
\label{qr}

 At low temperature 
 the static approximation is not justified and 
 we must use a different method.
 Introducing the auxiliary variables
 for the expression (\ref{trel}), 
 we write the last term in a linearized form 
\be
 & & 
 \exp\left[\frac{J^2}{2}\sum_{a=1}^n\int_{\tau,\tau'}
 \tilde{\chi}(\tau,\tau')n^a_{z}(\tau) n^a_{z}(\tau')
 \right]
 \no\\
 & & 
 = \int {\cal D}\tilde{z}\exp\left[
 -\frac{1}{2}\sum_a\int_{\tau,\tau'}\tilde{z}_a(\tau)
 \chi^{-1}(\tau-\tau')\tilde{z}_a(\tau')
 +J\sum_a\int_\tau \tilde{z}_a(\tau)n^a_z(\tau)
 \right]. 
 \label{chinn}
\ee
 As we analyzed in the Landau expansion 
 $\chi(\tau)$ decays exponentially in time 
 at $\tau=0$ and $\beta$.
 This decay rate is large at low temperature 
 and we can use the instantaneous approximation.
 Using the derivative expansion for $\tilde{z}(\tau)$, we obtain
\be
 \int {\cal D}\tilde{z}\exp\left[
 -\frac{1}{2}\sum_a\int_{\tau,\tau'}\tilde{z}_a(\tau)
 \chi^{-1}(\tau-\tau')\tilde{z}_a(\tau')
 \right]
 \sim \exp\left[
 -\frac{1}{2\beta\chi}\sum_a\int_\tau \tilde{z}_a^2(\tau)
 -\frac{\chi_2}{4\beta^2\chi^2}\sum_a\int_\tau 
 (\partial_\tau \tilde{z}_a)^2(\tau)
 \right],
\ee
 where
\be
 \chi=\frac{1}{\beta}\int_\tau\chi(\tau), \quad
 \chi_2=\frac{1}{\beta}\int_\tau\tau^2\chi(\tau).
\ee
 $\chi_2$ can be written as 
 $\chi_2=-\partial_\omega^2\chi(\omega)|_{\omega=0}$
 and we see that 
 this contribution was not taken into account 
 in the previous approximation.
 Assuming the form 
 $\chi(\tau)\sim e^{-2h'\tau}+e^{-2h'(\beta-\tau)}$ 
 we can write $\chi_2\sim (\beta\chi)^3/2$.
 The auxiliary variables are integrated out to give 
\be
 & & \int {\cal D}\tilde{z}\exp\left[
 -\frac{1}{2\beta\chi}\sum_a\int_\tau \tilde{z}_a^2(\tau)
 -\frac{\chi_2}{4\beta^2\chi^2}\sum_a\int_\tau 
 (\partial_\tau \tilde{z}_a)^2(\tau)
 +J\sum_a\int_\tau \tilde{z}_a(\tau)n^a_z(\tau)
 \right] 
 \no\\
 & & 
 = \exp\left[
 \frac{J^2}{2}\sum_a\int_\tau G(\tau-\tau')n_z^a(\tau)n_z^a(\tau')
 \right],
 \label{Gnn}
\ee
 where
\be
 G(\tau)=\sum_{n\ne 0} 
 \frac{\chi^{-1}}{(\pi n)^2+\chi^{-2}}
 e^{-2\pi i n\tau/\beta}. 
 \label{G}
\ee
 Compared to Eq.(\ref{chinn}) we find that $\tilde{\chi}(\tau)$ is 
 replaced by $G(\tau)$.
 Then the correlation function of $n_z^a(\tau)n_z^a(\tau')$
 in Eq.(\ref{Gnn}) 
 is calculated  by the gauge transformation 
 as $D_{h_a}(\tau-\tau')\Gamma^2/h_a^2$.
 Taking the limit $n\to 0$, 
 we obtain approximately
\be
 \beta f_{\rm eff} = \frac{\beta^2 J^2}{4}(\chi^2-q^2)
 -\int Dz_1\ln\left[\int Dz_2 (e^{\beta h}+e^{-\beta h})\right]
 -\frac{\beta^2J^2}{2}
 \int Dz_1D'z_2
 \frac{\Gamma^2}{h^2}\frac{\chi}{1+\beta h\chi}.
\ee
 This effective free energy is valid at low temperature and
 the phase transition point at $T=0$
 can be determined from 
\be
 & & \beta J\chi=\frac{1}{\beta J\chi}\left(\int D'z z^2-1\right)
 +\frac{\beta J}{2}\int D'z\left(z^2-\int D'z' z'^2\right)
 \frac{\Gamma^2}{h^2}\frac{1}{1+\beta h\chi}, \no\\
 & & 1=\frac{1}{(\beta J\chi)^2}\left(\int D'z z^2-1\right)^2.
\ee
 These equations have the solution 
 $\beta J\chi\sim 0.62$ and $\Gamma/J\sim 1.62$ at $T\to 0$.
 We see in Fig.\ref{spb} that the result of the quantum regime is 
 smoothly connected to that of the classical regime.

 Finally we determine the boundary between 
 the classical and quantum regime.
 According to Eq.(\ref{G}), the time dependence becomes important
 when the first Matsubara frequency is comparable to 
 the ``mass'' term of the propagator.
 Thus we identify the quantum regime 
 $\pi\chi\ $\raisebox{-.7ex}{$\stackrel{\textstyle <}{\sim}$}$\,\ 1$
 which is illustrated in Fig.\ref{spb}.

\section{Conclusions}
\label{conc}

 We have discussed quantum random spin systems using 
 the field theoretical method based on 
 the spin coherent state representation.
 The partition function is represented as a 
 functional integral and the continuous integration variables 
 describe the spin motion on a unit sphere.
 This formulation can be applied to arbitrary Hamiltonians 
 with arbitrary spin $S$.

 In previous works for the Ising spin ($S=1/2$) systems, 
 the eigenstates of $\hat{S}_3$ has been used for 
 the closure relation to be inserted into the Trotter decomposition.
 This formulation gives discrete integration variables and 
 the continuum limit in the time direction cannot be taken 
 since the time derivative for the discrete variables is ill defined.
 In this sense our formulation is natural and useful even for 
 the Ising systems.

 Our method is also useful for explicit calculations.
 We can use various field theoretical techniques 
 such as the semiclassical method
 and the gauge transformation.
 As an application we considered 
 the transverse SK model. 
 The classical effective free energy 
 renormalized by quantum fluctuation effects 
 are expressed in terms of order parameters
 and the saddle point equations are solved to obtain the phase diagram.
 We showed that the time dependence of $\chi$ is important 
 to obtain the result.
 We found the quantum phase transition point 
 located between the perturbative and the static results.
 Our estimate is $\Gamma/J= 1.62$ at $T=0$ 
 and is close to the values obtained by others.~\cite{YI,MH,RGAR}

 What is conceptually important in our calculation is that 
 the role of the order parameter variables is distinguished 
 between the static and nonstatic parts.
 The order parameter is defined as 
 the static part of the variable
 and the nonstatic part are integrated out 
 to find the effective classical theory,
 which is reminiscent of the renormalization group theory.
 Therefore, it is a straightforward extension 
 to consider the renormalization group calculation 
 as was discussed in Ref.\onlinecite{RSY}.
 Allowing for the spatial fluctuations of the order parameters,
 we can examine the stability of the critical states and 
 calculate the critical exponents.

 We clarified in the present paper 
 the role of the quantum fluctuations 
 using the simple transverse Ising spin glass model.
 It is a straightforward task to apply our results 
 to other models.
 For example we can consider the transverse SK model  
 with arbitrary spin $S$. 
 In that case, it is not difficult to calculate 
 the correlation function corresponding to Eq.(\ref{Ddef}). 
 When $S$ is large, we find that the time dependence becomes weak  
 and the static approximation becomes a good one.
 In a similar reason, the static approximation is 
 justified when we have an infinite many-body interaction.~\cite{re}
 These observations show that the present transverse Ising spin model 
 is the simplest one but the quantum effect is maximum.
 We think the next simplest nontrivial application of our method 
 is the quantum Heisenberg model.
 Most of the previous works relied on a semiclassical method 
 such as the static approximation \cite{BM,GL2}
 and the large-$N$ limit \cite{SYGPS}.
 We hope that our approach will be useful 
 for studying the quantum effects.

 Finally we mention another possible application. 
 In the present paper we considered the imaginary time 
 formulation to calculate the partition function.
 It is also possible to consider the real time formulation
 which allows us to analyze the dynamical correlations.
 This can be done by using the Keldysh formulation.~\cite{Keldysh}
 We can calculate the dynamical correlation function without 
 using the analytic continuation from imaginary to real time.
 The Keldysh method is also useful when we consider 
 the random averaging and field theoretical methods were developed 
 for disordered Fermion systems.~\cite{Kdiss}
 The application to the random quantum spin systems 
 is an interesting problem and is left for future work.

\section*{ACKNOWLEDGMENT}

 The author is grateful to 
 H. Nishimori and T. Obuchi 
 for useful discussions and comments.

\appendix*
\section{Derivation of the saddle point equations}

 We consider the derivative of the following functions
 to derive the saddle point equations:
\be
 F=\int Dz_1\ln\left[\int Dz_2 f(h)\right], \quad
 G=\int Dz_1\frac{\int Dz_2 f(h)g(h)}{\int Dz_2 f(h)},
\ee
 where
 $h=\left(\Gamma^2+M^2\right)^{1/2}$ with $M=J(\sqrt{q}z_1+\sqrt{\chi-q}z_2)$.
 After taking the derivative with respect to $\chi$ or $q$, 
 we take the limit $q=0$.
 First, we consider the derivative with respect to $\chi$.
 We have
\be
 \frac{\partial F}{\partial \chi}
 &=& \frac{J}{2}\int Dz_1\frac{\int Dz_2\frac{z_2}{\sqrt{\chi-q}}
 \frac{\partial}{\partial M}f(h)}{\int Dz_2 f(h)} \\
 &=& \frac{J}{2}\int Dz_1\frac{\int Dz_2
 \frac{1}{\sqrt{\chi-q}}
 \frac{\partial}{\partial z_2}
 \frac{\partial}{\partial M}f(h)}{\int Dz_2 f(h)} \\
 &=& \frac{J^2}{2}\int Dz_1\frac{\int Dz_2
 \frac{\partial^2}{\partial M^2}f(h)}{\int Dz_2 f(h)},
\ee
 where we referred the definition of the integration measure (\ref{Dz})
 to use $ze^{-z^2/2}=-de^{-z^2/2}/dz$
 and the partial integration in the second line.
 In the same way, we have
\be
 \frac{\partial G}{\partial \chi}
 = \frac{J^2}{2}\int Dz_1\left[
 \frac{\int Dz_2 \frac{\partial^2}{\partial M^2}f(h)g(h)}{\int Dz_2f(h)}
 -\frac{\int Dz_2 \frac{\partial^2}{\partial M^2}f(h)}{\int Dz_2f(h)}
 \frac{\int Dz_2 f(h)g(h)}{\int Dz_2f(h)}
 \right].
\ee
 At the limit $q=0$, $h$ becomes independent of $z_1$ and we obtain
\be
 \left.\frac{\partial F}{\partial \chi}\right|_{q=0}
 &=& \frac{1}{2\chi}\frac{\int Dz
 (z^2-1)f(h)}{\int Dz f(h)}, 
 \\
 \left.\frac{\partial G}{\partial \chi}\right|_{q=0}
 &=& \frac{1}{2\chi}\left[
 \frac{\int Dz (z^2-1)f(h)g(h)}{\int Dzf(h)}
 -\frac{\int Dz (z^2-1)f(h)}{\int Dzf(h)}
 \frac{\int Dz f(h)g(h)}{\int Dzf(h)}
 \right],
\ee
 where $M=J\sqrt{\chi} z$.
 We replaced the derivative with respect to $M$ by 
 that with respect to $z$ 
 and used again the partial integration.
 In the same way, we obtain 
\be
 \left.\frac{\partial F}{\partial q}\right|_{q=0} 
 &=& -\frac{1}{2\chi}\left[\frac{\int Dz 
 zf(h)}{\int Dz f(h)}\right]^2, 
 \\
 \left.\frac{\partial^2 F}{\partial q^2}\right|_{q=0} &=& 
 -\frac{1}{2\chi^2}\left[
 \frac{\int Dz (z^2-1)f(h)}{\int Dz f(h)}\right]^2,
 \\
 \left.\frac{\partial G}{\partial q}\right|_{q=0} &=& 0, 
 \\
 \left.\frac{\partial^2 G}{\partial q^2}\right|_{q=0}
 &=& -\frac{1}{\chi^2}
 \frac{\int Dz (z^2-1)f(h)}{\int Dz f(h)}
 \left[
 \frac{\int Dz z^2f(h)g(h)}{\int Dz f(h)}
 -\frac{\int Dz z^2f(h)}{\int Dzf(h)}
 \frac{\int Dz f(h)g(h)}{\int Dzf(h)}
 \right].
\ee



\begin{thebibliography}{99}
\bibitem{MPV}
M. M\'ezard, G. Parisi, and M.A. Virasoro, 
{\it Spin Glass Theory and Beyond}
(World Scientific, Singapore, 1987).

\bibitem{Nishimori}
H. Nishimori, {\it Statistical Physics of Spin Glasses and 
Information Processing: An Introduction} 
(Oxford University Press, Oxford, 2001).

\bibitem{EA}
S.F. Edwards and P.W. Anderson, J. Phys. F: Met. Phys. {\bf 5}, 965 (1975).

\bibitem{SK} 
D. Sherrington and S. Kirkpatrick, 
Phys. Rev. Lett. {\bf 35}, 1792 (1975).

\bibitem{CDS}
B.S. Chakrabarti, A. Dutta, and P. Sen, 
{\it Quantum Ising Phases and Transitions in Transverse Ising Models}
(Springer, Berlin, 1996).

\bibitem{BM}
A.J. Bray and M.A. Moore, J. Phys. C {\bf 13}, L655 (1980).

\bibitem{Sachdev}
See for example, 
S. Sachdev, {\it Quantum Phase Transitions} 
(Cambridge University Press, Cambridge, 1999).

\bibitem{IY}
H. Ishii and T. Yamamoto, J. Phys. C {\bf 18}, 6225 (1986).

\bibitem{U}
K.D. Usadel, Solid State Commun. {\bf 58}, 629 (1986).

\bibitem{YI}
T. Yamamoto and H. Ishii, J. Phys. C {\bf 20}, 6053 (1987).

\bibitem{WL}
K. Walasek and K. Lukierska-Walasek, 
Phys. Rev. B {\bf 38}, 725 (1988).

\bibitem{Y}
T. Yokota, Phys. Rev. B {\bf 40}, 9321 (1989).

\bibitem{TLK}
D. Thirumalai, Q. Li, and T.R. Kirkpatrick, J. Phys. A {\bf 22}, 3339 (1989).

\bibitem{GL}
Y.Y. Goldschmidt and P.Y. Lai, Phys. Rev. Lett. {\bf 64}, 2467 (1990).

\bibitem{UBK}
K.D. Usadel, G. B\"uttner, and T.K. Kope\'c, 
Phys. Rev. B {\bf 44}, 12583 (1991).

\bibitem{MH}
J. Miller and D.A. Huse, Phys. Rev. Lett. {\bf 70}, 3147 (1993).

\bibitem{RGAR}
M.J. Rozenberg and D.R. Grempel, Phys. Rev. Lett. {\bf 81}, 2550 (1998); 
L. Arrachea and M.J. Rozenberg, {\it ibid.} {\bf 86}, 5172 (2001).

\bibitem{KK}
D.H. Kim and J.J. Kim, Phys. Rev. B {\bf 66}, 054432 (2002).

\bibitem{TS}
H.F. Trotter, Proc. Am. Math. Soc. {\bf 10}, 545 (1959); 
M. Suzuki, Prog. Theor. Phys. {\bf 56}, 1454 (1976).

\bibitem{SYGPS}
S. Sachdev and J. Ye, Phys. Rev. Lett. {\bf 70}, 3339 (1993); 
A. Georges, O. Parcollet, and S. Sachdev, 
{\it ibid.} {\bf 85}, 840 (2000). 

\bibitem{K}
J.R. Klauder, Phys. Rev. D {\bf 19}, 2349 (1979).

\bibitem{cs}
J.R. Klauder and B-S. Skagerstam, {\it Coherent States} 
(World Scientific, Singapore, 1985).

\bibitem{WFS}
P.B. Wiegmann, Phys. Rev. Lett. {\bf 60}, 821 (1988);
E. Fradkin and M. Stone, Phys. Rev. B {\bf 38}, 7215 (1988).

\bibitem{AG}
A. Alscher and H. Grabert, J. Phys. A {\bf 32}, 4907 (1999).

\bibitem{SF}
M. Stone, Phys. Rev. D {\bf 33}, 1191 (1986); 
K. Fujikawa, {\it ibid.} {\bf 73}, 025017 (2006).

\bibitem{RSY}
J. Ye, S. Sachdev, and N. Read, Phys. Rev. Lett. {\bf 70}, 4011 (1993);
N. Read, S. Sachdev, and J. Ye, Phys. Rev. B {\bf 52}, 384 (1995).

\bibitem{Parisi}
G. Parisi, 
Phys. Rev. Lett. {\bf 43}, 1754 (1979); 
J. Phys. A {\bf 13}, L115 (1980); {\bf 13}, L1101 (1980); {\bf 13}, L1887 (1980).

\bibitem{re}
Y.Y. Goldschmidt, Phys. Rev. B {\bf 41}, 4858 (1990);
T. Obuchi, H. Nishimori, and D. Sherrington, 
J. Phys. Soc. Jpn. {\bf 75}, 054002 (2007).

\bibitem{GL2}
Y.Y. Goldschmidt and P.Y. Lai, Phys. Rev. B {\bf 43}, 11434 (1991).

\bibitem{Keldysh}
L.V. Keldysh, Sov. Phys. JETP {\bf 20}, 1018 (1965);
See for Fermion systems, 
J. Rammer and H. Smith, Rev. Mod. Phys. {\bf 58}, 323 (1986);
For spin systems, 
M.N. Kiselev and R. Oppermann, Phys. Rev. Lett. {\bf 85}, 5631 (2000).

\bibitem{Kdiss}
A. Kamenev and A.V. Andreev, 
Phys. Rev. B {\bf 60}, 2218 (1999); 
C. Chamon, A.W.W. Ludwig, and C. Nayak, 
{\it ibid.} {\bf 60}, 2239 (1999).

\end{thebibliography}
\end{document}